\documentclass[journal]{IEEEtran}

\usepackage{subfigure}
\usepackage{graphicx}
\usepackage{color}
\usepackage{placeins}
\usepackage{float}
\usepackage{tabularx,colortbl}
\usepackage{pbox}

\usepackage{subfigure}
\usepackage{cite}
\usepackage[cmex10]{amsmath}
\usepackage{amsfonts}
\usepackage{subscript}
\usepackage{epstopdf}

\usepackage{bm}
\usepackage{upgreek}
\makeatletter
\newcommand{\bfgreek}[1]{\bm{\@nameuse{up#1}}}
\makeatother
\begin{document}
\title{Origami Inspired Reconfigurable Antenna for Wireless Communication Systems}

\author{Ali~Molaei, Chang~Liu, Samuel M.~Felton, and~Jose~Martinez-Lorenzo% <-this % stops a space
\thanks{A. Molaei, Chang Liu, Samuel M. Felton, and Jose Martinez are with the College of Engineering, Northeastern University, Boston, MA, 02115 USA (e-mail: j.martinez-lorenzo@northeastern.edu).}% <-this % stops a space
}

\maketitle

\begin{abstract}
%\boldmath
This paper presents the design, fabrication, and experimental validation of an origami-inspired reconfigurable antenna. The proposed antenna can operate as a monopole or an inverted-L antenna, by changing its configuration. Doing so changes its operational frequency, principal radiation mode, and directivity. Measurements show that the antenna is able to change its resonance frequency from 750 MHz to 920 MHz---equivalent to 22.6\% frequency shift. Simulations are carried out to compare the radiation characteristics of the antenna (gain and radiation pattern) in both configurations. The results validate that the antenna has a reconfigurable bandwidth, which enables a better channel characterization for RF systems.
\end{abstract}



\IEEEpeerreviewmaketitle

\section{Introduction}

Origami engineering transforms flat materials into three dimensional structures with complex geometries. It has been widely used in various fields including self-assembly \cite{liu2017self} and mechanical design \cite{vander2014origamibot}. Origami-inspired structures have been used in many antenna engineering applications such as, circularly polarized \cite{liu2015origami,liu2015reconfigurable,yao2016polarization,shah2017low} and reconfigurable \cite{liu2014reconfigurable,yao2014novel,yao2014novel2,yao2015mode} antennas. Compared to the traditional antennas, origami-inspired antennas have many advantages for applications demanding mobility and accessible deployment.

%In the work we are presenting in this paper, we utilize an origami structure to provide different geometries, use them as antenna and study their behavior.

%Reconfigurable antennas play an important role in the modern wireless communication technologies.

In this work, a reconfigurable origami-based antenna has been presented. The proposed antenna operates as a monopole or an inverted-L antenna by switching its configuration. The radiation characteristics of the antenna in its two compositions has been studied in terms of return loss, radiation pattern, and realized gain. The presented antenna brings many interesting advantages to wireless communication networks. For instance, when the antenna is in its monopole mode, it would have a broadside radiation pattern and can be used for terrestrial communication applications. In addition, when the antenna is in its inverted-L mode, it would have an end-fire radiation pattern and can be used for satellite applications. The antenna is built using three layer laminating technique and is metallized using a silver spray, which makes it low-cost and light weight. The antenna could be easily reconfigured from monopole mode to inverted-L mode with a mechanical force.

\section{Design and Fabrication}

\subsection{Design}

Monopole and inverted-L antennas \cite{zhang2017antenna} and many modifications to them \cite{oh2004dual,lan2005design,ammann2003wideband} have been widely used in wireless communication applications. The monopole antenna has a principal mode in the broadside direction; the inverted-L antenna, however, has two principal modes in the broadside and end-fire directions. Nevertheless, the end-fire mode is stronger than the broadside mode, as the longer arm of the antenna is parallel to its ground plane. In this section, we will present the design of an origami-inspired antenna that can be reconfigured into two different compositions. In one composition it forms a monopole antenna and in the other one it forms an inverted-L antenna. Three main radiation characteristics of the antenna will be tailored by reconfiguring the antenna: the (1) resonance frequency, (2) principal radiation mode, and (3) directivity of the antenna. Since the inverted-L antenna leverages on the ground plane, it has a higher resonance frequency compared to the monopole antenna \cite{zhang2017antenna} --- this results in a frequency shift when the antenna is reconfigured from one state to the other one. In addition, the inverted-L antenna has a more directive beam than the monopole antenna \cite{zhang2017antenna}. Based on the radiation requirements of the targeted wireless communication application, the antenna can be reconfigured to have a broader beam-width and a lower gain, or to have a narrower beam-width and a higher gain. 

	\begin{figure}[!b]
		\centering
		\includegraphics[width=3.25in]{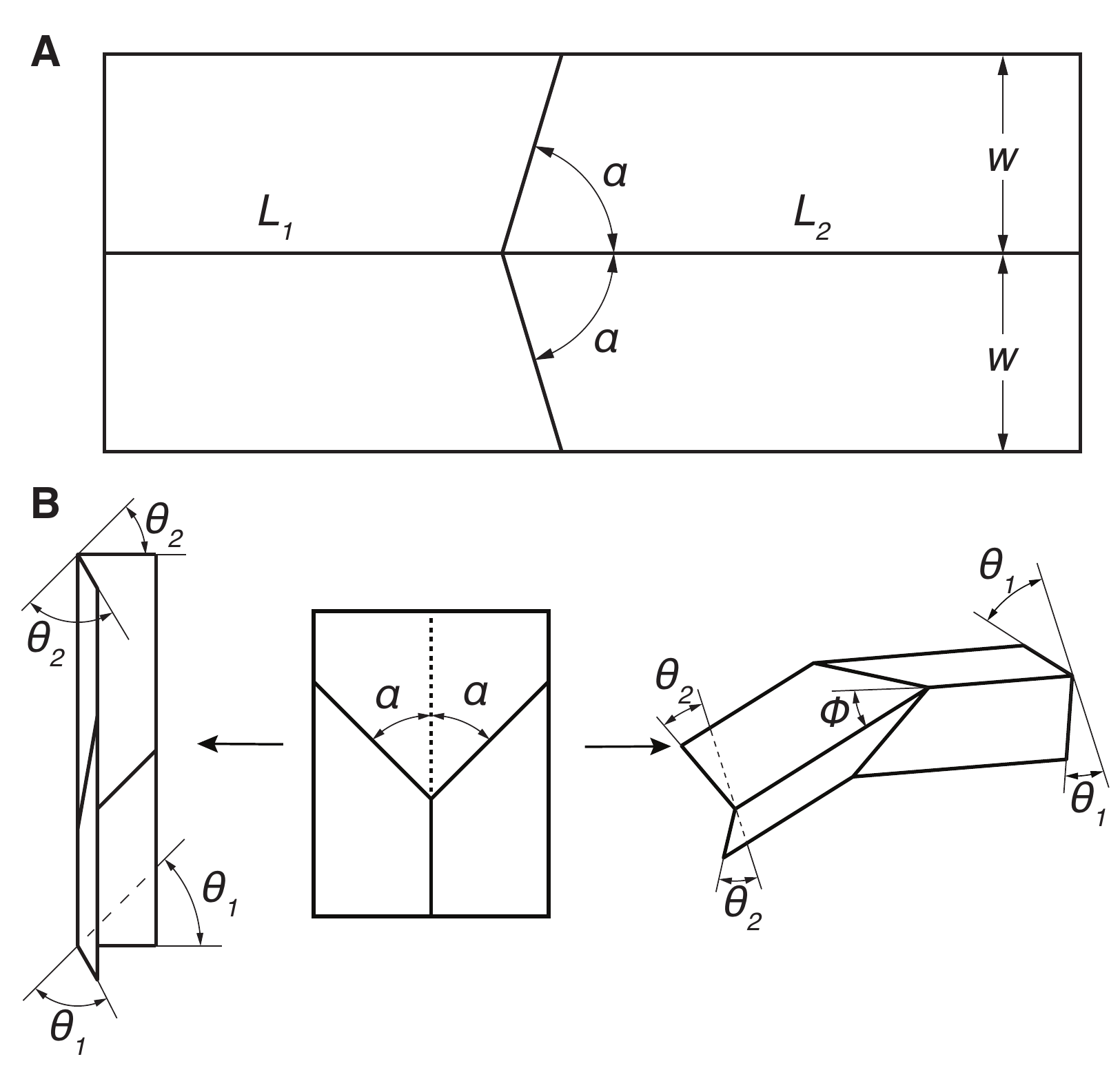}
		\caption{A, Muria origami pattern. B, Muria origami configurations. Left: parallel configuration; Middle: flat-unfolded configuration; Right: anti-parallel configuration.}
		\label{fig: Origami}
	\end{figure}

	The origami pattern used here is the Miura vertex. It is made out of four right-angle trapezoidal panels and connected by four creases (Fig. \ref{fig: Origami}). The two parallel creases are called spinal creases, and the other two are called peripheral creases \cite{zuliani2018minimally}. We define $\theta_{1}$ as the folding angle for the bottom section, $\theta_{2}$ as the the folding angle for the top section, and $\phi$ as angle between the spinal creases. For our sample, we chose the following parameters: $\alpha = 22.5 ^{\circ}$, $L_{1} =$ 25.4 mm, $L_{2} =$ 63.5 mm, and $w = $12.7 mm.
	
\begin{figure}[ht!]
    \centering
        \includegraphics[width=0.5\textwidth,trim={0cm 0cm 0.5cm 3cm},clip=true]{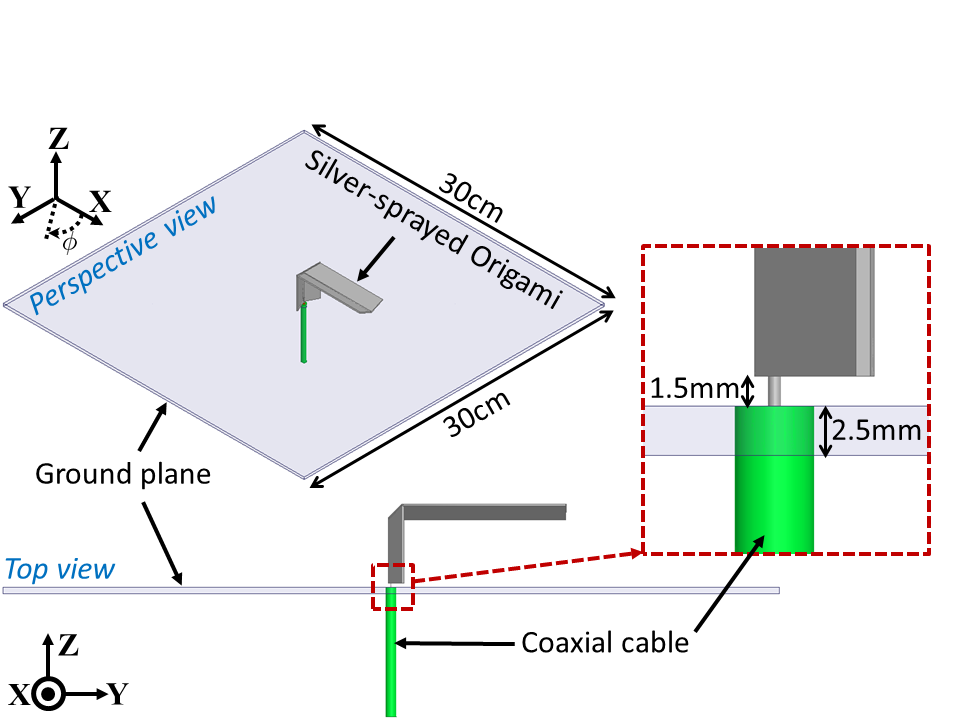} %\vspace{-0.3cm} 	
    \caption{ \label{Geometry} Geometry of the proposed reconfigurable antenna in the anti-parallel configuration.} %\vspace{-0.3cm}
\end{figure}

	This pattern has a single degree of freedom but can switch between two distinct configurations from the flat-unfolded state, each with their own shape and kinematic relationships (Fig. \ref{fig: Origami}B). We define that the Miura vertex is in the parallel configuration when $\textrm{sgn}(\theta_{1}) = \textrm{sgn}(\theta_{2})$ and $\phi = 0 ^{\circ}$ (Fig. \ref{fig: Origami}B(Left)), and in the anti-parallel configuration when $\textrm{sgn}(\theta_{1}) = -\textrm{sgn}(\theta_{2})$ and $\phi \neq 0 ^{\circ}$ (Fig. \ref{fig: Origami}B(Right)). When the parallel and anti-parallel configurations are mounted on ground plane, they operate as monopole and inverted-L antennas, respectively. Figure \ref{Geometry} illustrates the geometry of the origami pattern --- in the anti-parallel configuration --- integrated with a square metal plate (ground plane) and a coaxial probe to feed the antenna.

%ANSYS High-Frequency Structure Simulator (HFSS) has been used to design and fabricate the reconfigurable antennas.

\subsection{Fabrication}

	The antenna was fabricated using techniques established in previous papers on origami engineering \cite{felton2013self,zuliani2018minimally}. The origami vertex was laminated using three layers: one rigid structural layer made out of 800 $\upmu$m thick polystyrene sheet, one adhesive layer using double sided tape and one flexural layer made out of 50 $\upmu$m thick nylon film. Flexural hinges were created by removing the rigid polystyrene in thin lines, leaving only the nylon film to act as a flexural hinge.
	The polystyrene was machined with a CO$_{2}$ laser cutter (Universal Laser Systems, PLS6MW) to create gaps where the hinges would be placed. Then nylon film was taped to the surface of the cut polystyrene sheet using high-strength glue-on-a-roll (3M F9460PC). Finally, the assembled laminate piece was laser machined with a release cut around the mechanism edge.
	
\begin{figure}[ht!]
    \centering
        \subfigure[]{\label{fab-a}
        \includegraphics[width=0.4\textwidth,trim={0cm 0cm 7cm 7cm},clip=true]{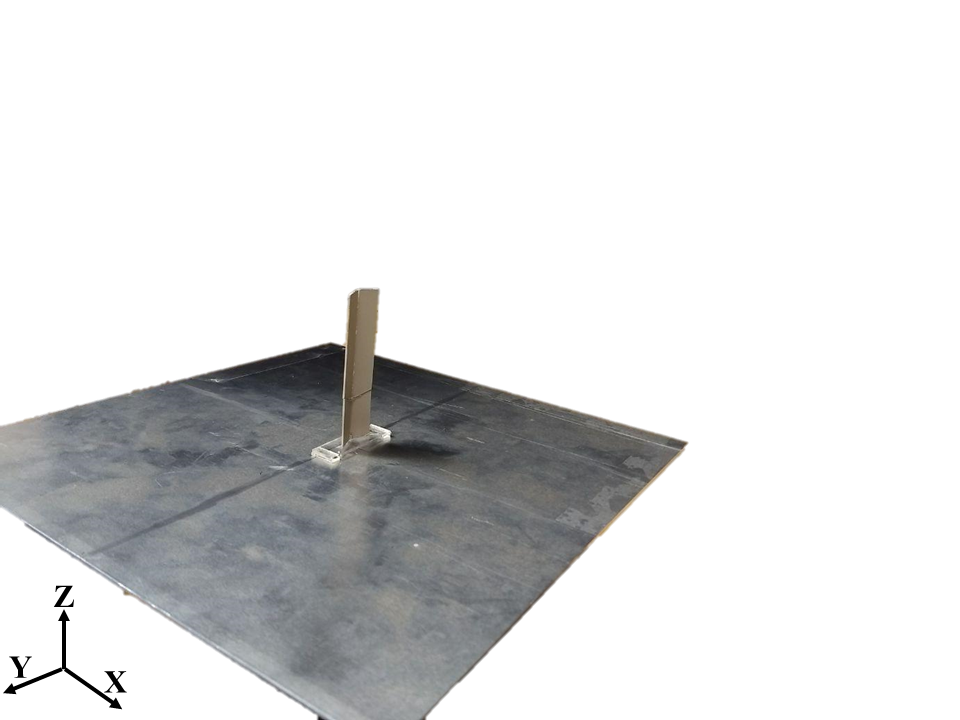}} %\hspace{-0.2cm}
				~
				\subfigure[]{\label{fab-b}
        \includegraphics[width=0.4\textwidth,trim={0cm 0cm 7cm 7cm},clip=true]{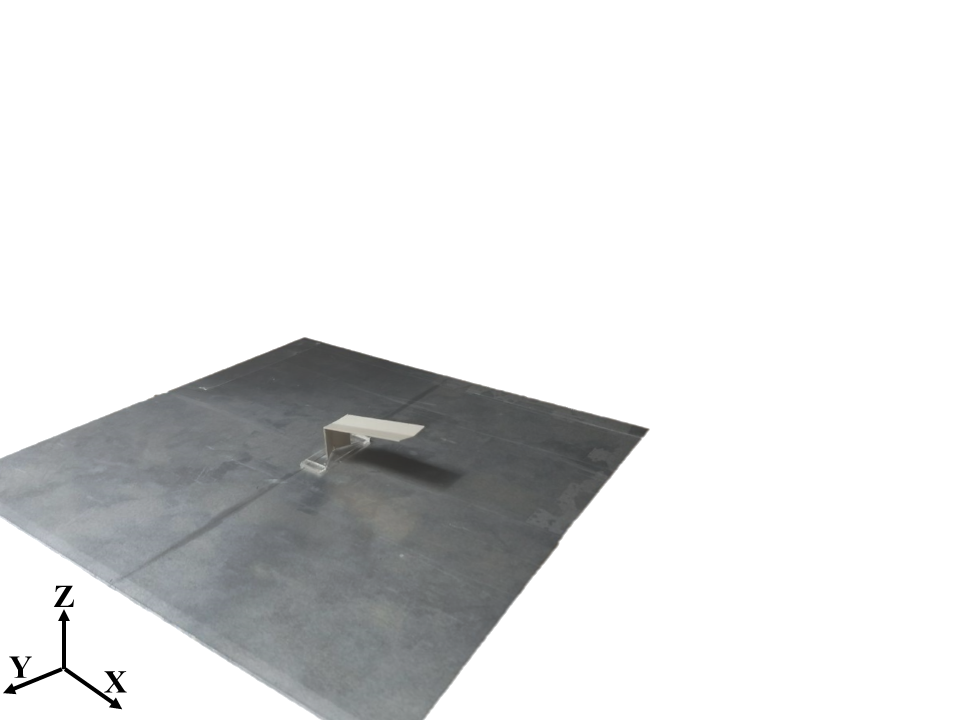}}  %\vspace{-0.3cm}	
    \caption{ \label{fab} Fabricated (a) parallel and (b) anti-parallel reconfigurable antenna.}  %\vspace{-0.2cm}
\end{figure}

	The assembled part was metalized using an acrylic-based silver conductive coating spray (842AR--SUPER SHIELD $^{TM}$). The part was sprayed two times on front and back with 10 minutes break between each spray to avoid trapping solvents between the coats. Then, the part was cured at room temperature for one hour. The antenna part was then mounted on top of a 30 cm$\times$30 cm metal plate (ground plane) with a 1.5 mm spacing between the antenna and the metal plate. Ultimately, a 50-ohm coaxial probe is connected to the base of the origami part and the ground plane to feed the antenna. Figures \ref{fab-a} and \ref{fab-b}, respectively, show the perspective view of the fabricated antenna in parallel/monopole and anti-parallel/inverted-L configurations.

\section{Simulation \& Experimental Results}

\begin{figure}[ht!]
    \centering
        \includegraphics[width=0.5\textwidth,trim={0cm 0cm 3cm 5cm},clip=true]{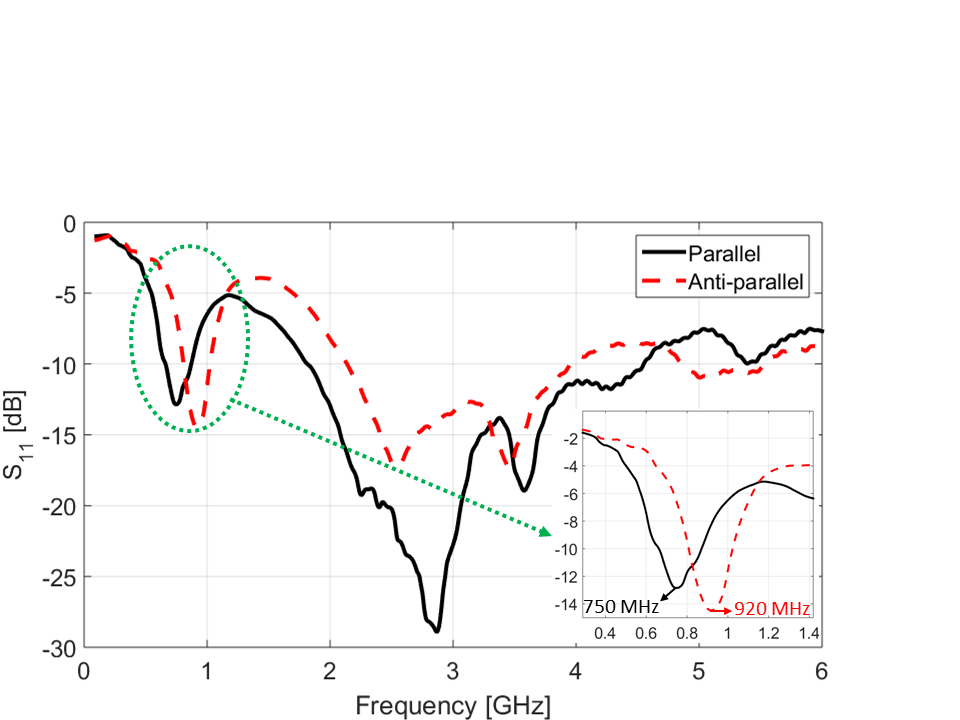} %\vspace{-0.3cm} 	
    \caption{ \label{RL} Measured $S_{11}$ for the fabricated parallel and anti-parallel reconfigurable antennas.} %\vspace{-0.3cm}
\end{figure}

Figure \ref{RL} shows the $|S_{11}|$ of the fabricated antenna for both parallel and anti-parallel configurations. The measurement is done by an N5242A PNA-X Network Analyzer calibrated in the frequency range of 0.2 GHz to 6 GHz. As can be seen in Fig. \ref{RL}, the principal/first resonance of the antenna shifts from 750 MHz to 920 MHz, when reconfiguring the antenna from the parallel to the anti-parallel arrangement --- corresponding to a 22.6\% frequency shift. The bandwidth (BW) of the parallel and anti-parallel configurations are measured to be 28.3\% and 23.7\%, respectively --- expressing that the BW of the antenna is slightly affected --- when the antenna is reconfigured from one state to the other one. This is of interest for wireless communication applications with BW stability requirements.

\begin{figure}[ht!]
    \centering
        \subfigure[]{\label{EP-a}
        \includegraphics[width=0.22\textwidth,trim={0cm 0cm 4.4cm 0cm},clip=true]{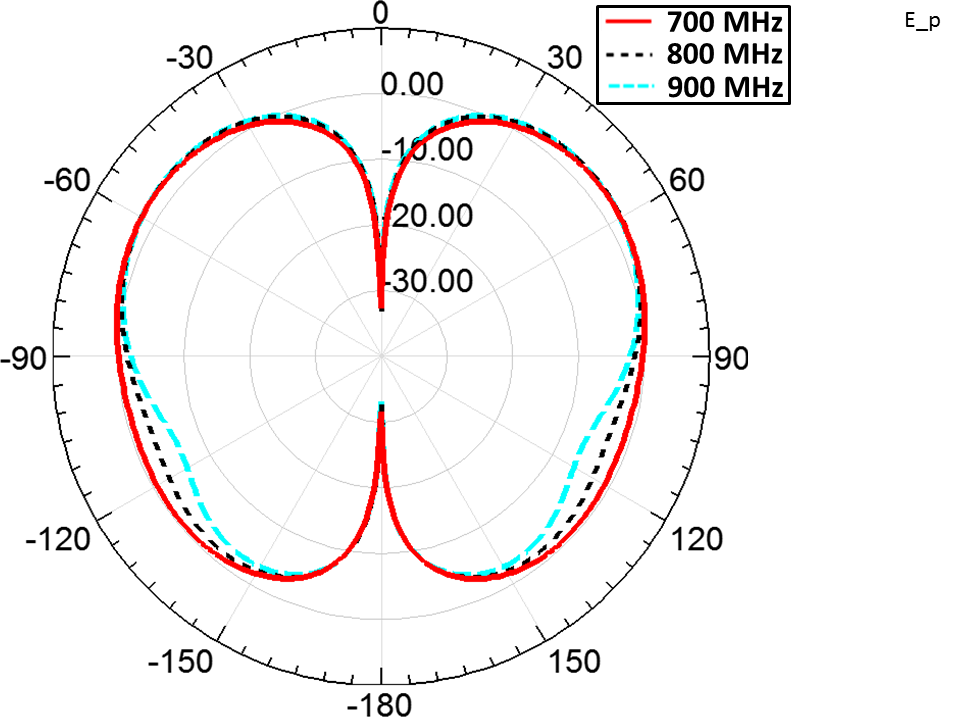}} %\hspace{-0.2cm}
				~
				\subfigure[]{\label{EP-b}
        \includegraphics[width=0.22\textwidth,trim={0cm 0cm 4.4cm 0cm},clip=true]{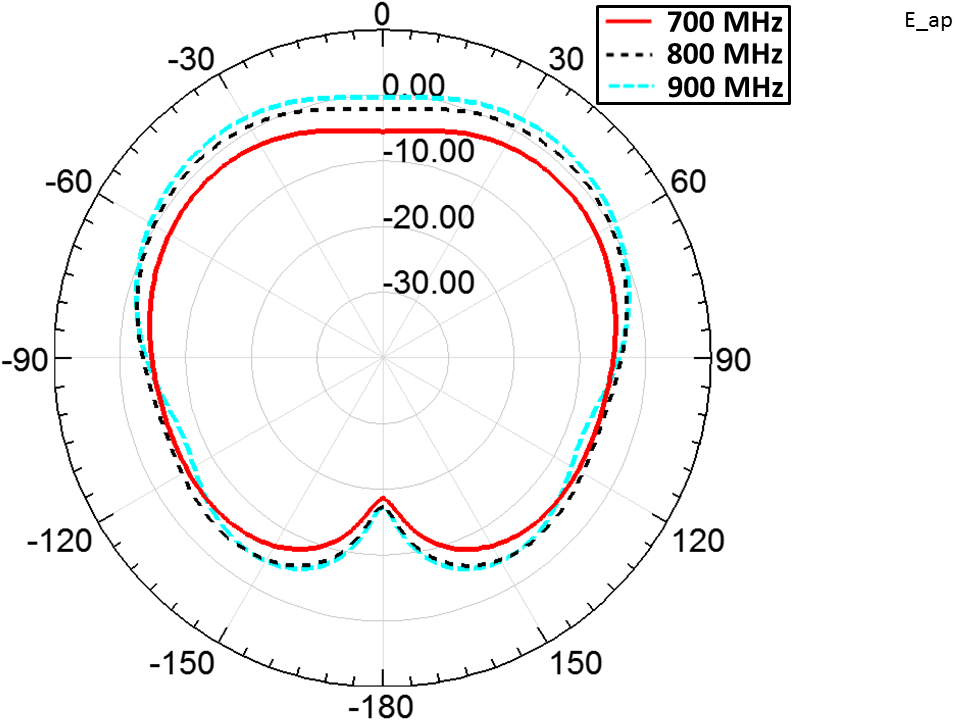}}  %\vspace{-0.3cm}	
    \caption{ \label{EP} E-Plane realized gain pattern of the (a) parallel and (b) anti-parallel reconfigurable antennas.}  %\vspace{-0.2cm}
\end{figure}

\begin{figure}[ht!]
    \centering
        \subfigure[]{\label{HP-a}
        \includegraphics[width=0.22\textwidth,trim={0cm 0cm 4.4cm 0cm},clip=true]{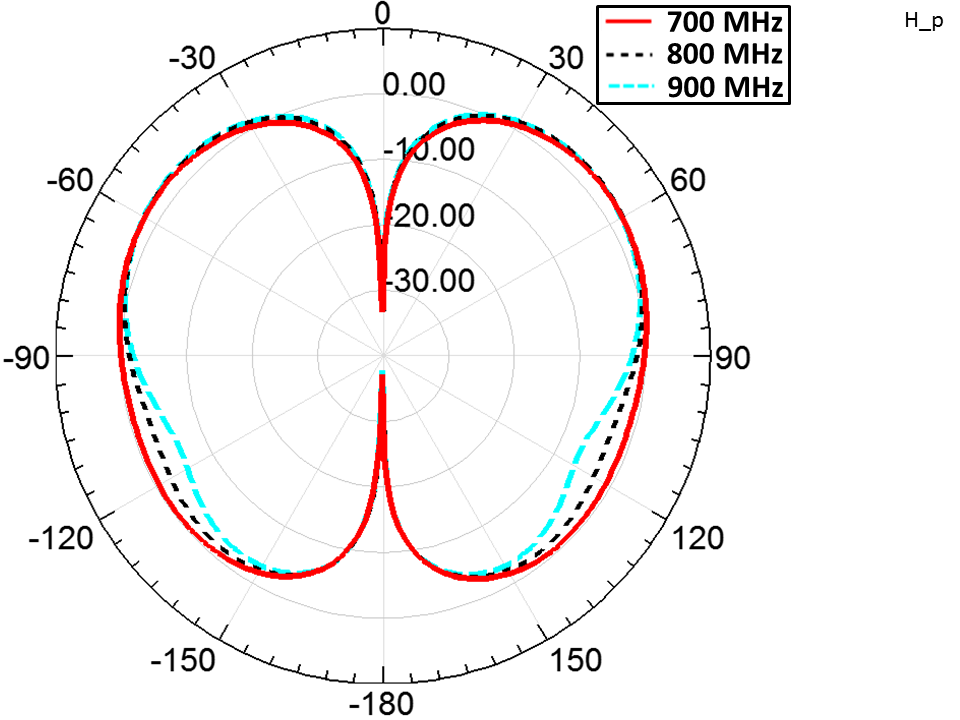}} %\hspace{-0.2cm}
				~
				\subfigure[]{\label{HP-b}
        \includegraphics[width=0.22\textwidth,trim={0cm 0cm 4.4cm 0cm},clip=true]{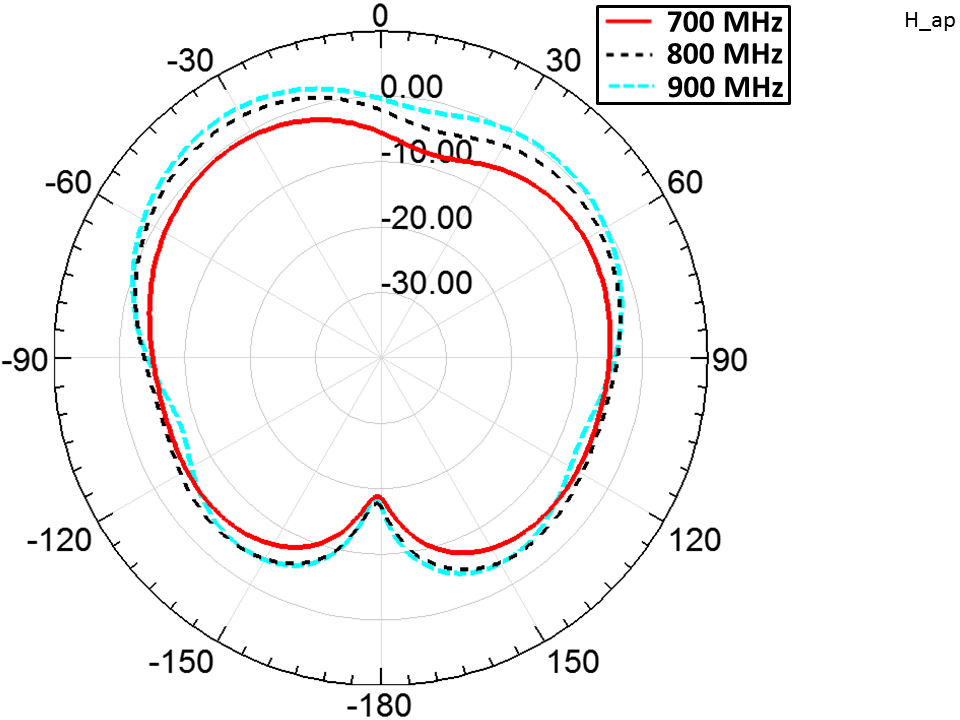}}  %\vspace{-0.3cm}	
    \caption{ \label{HP} H-Plane realized gain pattern of the (a) parallel and (b) anti-parallel reconfigurable antennas.}  %\vspace{-0.2cm}
\end{figure}

\begin{figure}[ht!]
    \centering
        \subfigure[]{\label{3D-a}
        \includegraphics[width=0.22\textwidth,trim={0cm 0cm 6cm 5.3cm},clip=true]{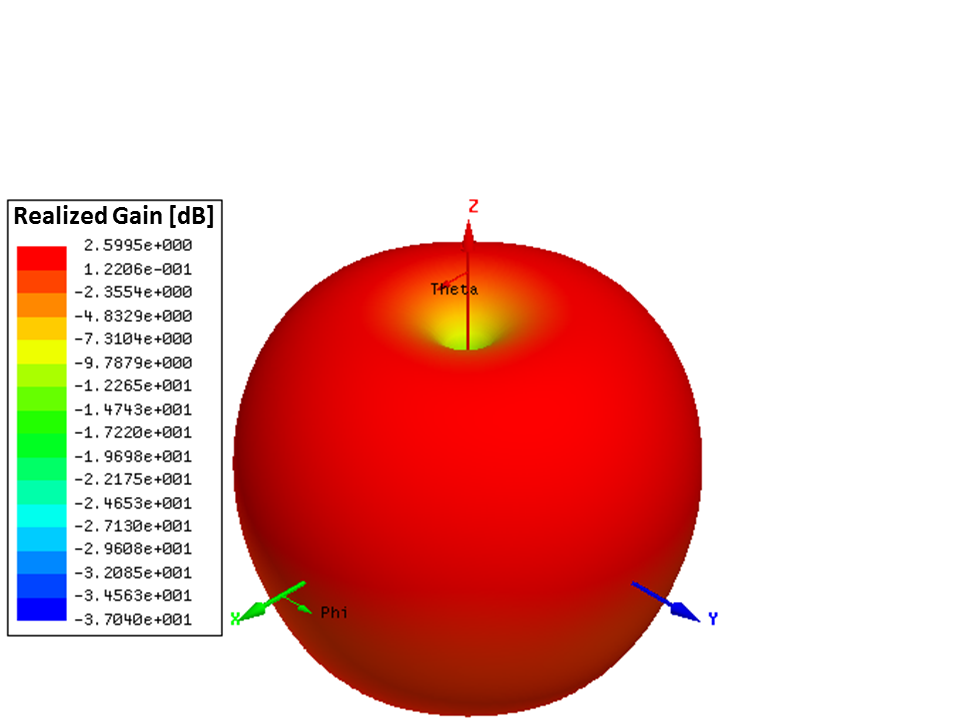}} %\hspace{-0.2cm}
				~
				\subfigure[]{\label{3D-b}
        \includegraphics[width=0.22\textwidth,trim={0cm 0cm 6cm 5.3cm},clip=true]{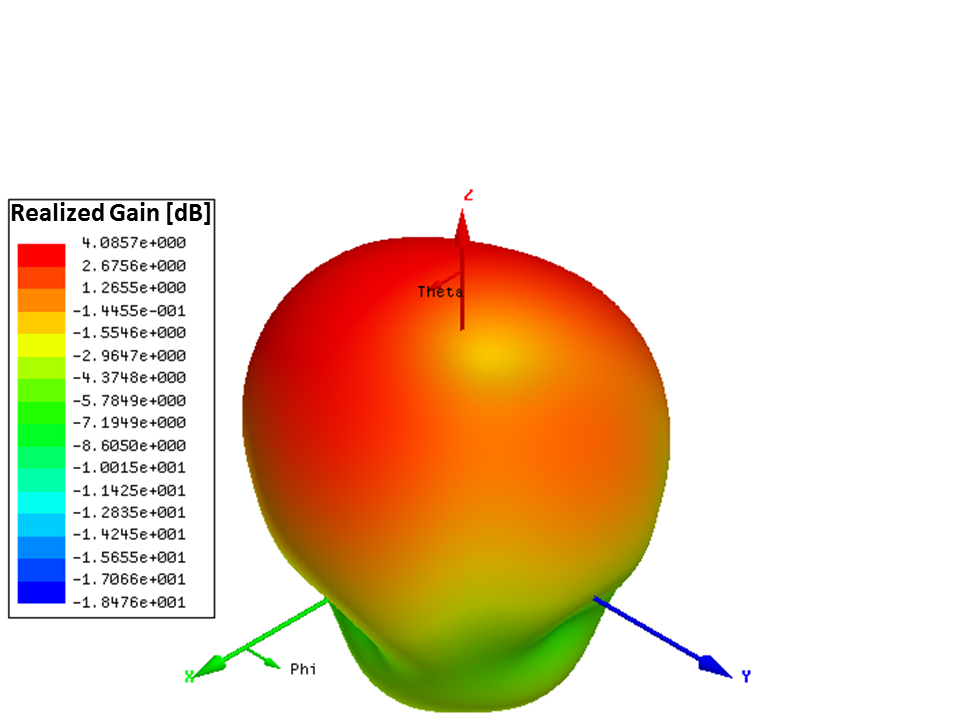}}  %\vspace{-0.3cm}	
    \caption{ \label{3D} 3D realized gain pattern of the (a) parallel configuration at 750 MHz and (b) anti-parallel configuration at 920 MHz.}  %\vspace{-0.2cm}
\end{figure}

The E-plane ($\phi=90^{\circ}$) and H-plane ($\phi=0^{\circ}$) are defined as the two fundamental far-field planes for studying the radiation pattern of the antenna. Figures \ref{EP} and \ref{HP} represent the realized gain radiation pattern of the antenna in its two principal configurations in the E-plane and H-plane, respectively. As can be seen in Fig. \ref{HP-b}, in the anti-parallel configuration, the peak of the radiation pattern in H-plane is not exactly aligned in the end-fire direction. This is due to the asymmetric geometry of the antenna with respect to the YZ-plane. Figure \ref{3D} shows the far-field realized gain pattern of the antenna at the resonance frequencies for its two configurations --- 750 MHz for the parallel and 920 MHz for the anti-parallel configurations.

\begin{figure}[ht!]
    \centering
        \includegraphics[width=0.5\textwidth,trim={0cm 0cm 3cm 5cm},clip=true]{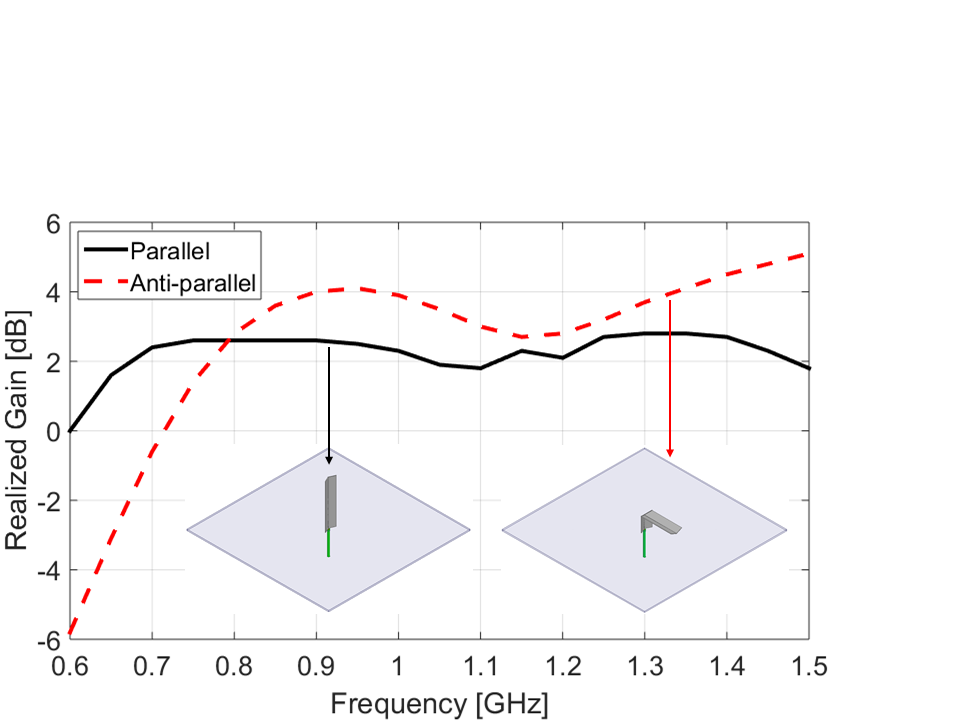} %\vspace{-0.3cm} 	
    \caption{ \label{Gain} Realized gain of the simulated parallel and anti-parallel reconfigurable antennas.} %\vspace{-0.3cm}
\end{figure}

The realized gain of the simulated antennas are calculated at the direction in which radiated power is maximum (shown in Fig. \ref{Gain}). As can be seen, the anti-parallel configuration has a higher gain for frequencies larger than 800 MHz, verifying its more directive radiation, compared to that of the parallel configuration. Radiation performance of the two configurations is summarized in Table \ref{Table1}.

\begin{table}[ht!]
\centering \caption{Design parameters of the CHA and SPPs.}\label{Table1} %\setlength{\extrarowheight}{1.5pt}
\begin{tabular}{|l|l|l|l|}
\hline
{\textbf{Conf.}} & {\textbf{Res. freq. ($f_r$)}}   & {\textbf{HPBW}}     & {\textbf{Gain at $f_r$}} \\
\hline
{Parallel}               & {750 MHz}               & {$66^{\circ}$}      & {2.6 dB}\\
\hline
{Anti-parallel}          & {920 MHz}               & {$57^{\circ}$}      & {4 dB}\\
\hline 
\end{tabular}
\end{table}

\section{Conclusion}
This letter presents the preliminary design, fabrication, and experimental validation of an origami-inspired reconfigurable antenna. It is shown that the proposed antenna is reconfigurable, i.e. it can change its operational frequency, direction of maximum radiation pattern, and maximum realized gain, based on its configuration. In this letter, the origami structure was constructed to comply with the original designs of the monopole and inverted-L antennas. Nevertheless, additional studies will be performed in order to explore the use of matching networks, as modifiers to the original designs; this will enhance the matching between the coaxial cable and the antenna structure, thus leading to improvement in the gain and BW of the antenna.

%\section*{ACKNOWLEDGEMENT}
%\textcolor[rgb]{1,0,0}{This work has been funded by ???}.

\bibliography{Origami_Ant,references}
\bibliographystyle{IEEEtran}

\end{document}